\documentstyle[12pt]{article}
\begin{document}
\baselineskip=1.\baselineskip

\makebox[7cm]

\vskip 1cm

\vspace*{1cm}
\begin{center}
\begin{Large}
\begin{bf}
 NNLO QCD Analysis  of CCFR data on $xF_3$ Structure Function
and Gross-Llewellyn Smith Sum Rule\\
with Higher Twist and Nuclear Corrections
\\
\end{bf}
\end{Large}

\vspace*{15mm}
{\large
M.V. Tokarev$^{b}$,\footnote{E-mail: tokarev@sunhe.jinr.dubna.su}
\ \ \ A.V. Sidorov$^a$\footnote{E-mail: sidorov@thsun1.jinr.dubna.su}
}\\


\vskip 1.5cm

$^a$Laboratory of High Energies,\\
Joint Institute for Nuclear Research,\\
141980 Dubna, Moscow Region, Russia\\

\vskip 0.2cm

$^b$Bogoliubov Laboratory of Theoretical Physics,\\
Joint Institute for Nuclear Research,\\
141980 Dubna, Moscow Region, Russia\\

\end{center}
\vskip 1cm

\begin{abstract}
A detailed NNLO QCD  analysis
of new CCFR  data
on the $xF_3$  structure function
including  the target  mass, higher twist and nuclear
corrections
was performed and parametrizations of the perturbative and power
terms  of the structure function were constructed.
The results of  QCD analysis of the structure function were used to
study the $Q^2-$dependence of the Gross-Llewellyn Smith sum rule.
The $\alpha_S/\pi-$expansion  of $S_{GLS}(Q^2)$  was studied
and parameters of the expansion  were found to be
$s_1=2.74\pm 0.01$, $s_2=-2.22\pm 0.23$, $s_3=-7.86\pm 1.74$
in a good agreement with the perturbative QCD predictions  for the
Gross-Llewellyn Smith sum rule in the next-to-leading
 and next-to-next-to-leading order. \\ [5mm]
\end{abstract}

~~~~~~~~PACS: 12.38.Qk; 12.38.Bx; 13.15.Em. \\[8mm]

\newpage

{\section {Introduction}}

The progress of perturbative Quantum ChromoDynamics (QCD) in the description
of the high energy physics of strong interactions is considerable \cite{QCD20}.
Recent experimental data on the structure function of neutrino
deep-inelastic scattering obtained at the Fermilab Tevatron \cite{CCFR97}
provide a good possibility to precisely verify the QCD predictions for
scaling violation with experiment  data.
The QCD predictions for the structure functions evolution of
deep-inelastic scattering (DIS) are calculated now up to the
next--to--next--to--leading order (NNLO) of a perturbative theory
\cite{VZ1,LV,VZ2,LRV}.
The method of comparison of 3--loop QCD predictions
with SF experimental data has been developed in \cite{PKK,KKPS1,KKPS2} based
on the Jacobi polynomial structure function expansion \cite{PaSu}.
It is well known that beyond the perturbative QCD
 there are other effects (higher twist effects,
nuclear corrections, target mass corrections etc.) to be included
into the joint QCD analysis of SF.

In the paper, we present the results of  NNLO QCD  analysis
of the data on the $xF_3$  structure function  obtained by the CCFR
 Collaboration
with taking in account of the target  mass, higher twist and nuclear
  corrections. The $Q^2$--evolution of SF is studied
and the parametrizations of perturbative (leading twist) and power
terms  of the structure function are  constructed.
The results of our NNLO QCD analysis of the structure function  are in
good qualitative agreement with NNLO theoretical predictions
for the $Q^2$-evolution of the Gross-Llewellyn Smith sum rule \cite{GLS}
$S_{GLS}^{\it theor}(Q^2)=
3\cdot [1 -\alpha_S/\pi- 3.25\cdot (\alpha_S/\pi)^2]$ \cite{gls3loop}.

In Section 2, the method of NNLO QCD analysis of SF based on
the SF Jacobi polynomial expansion including the
target  mass, higher twist and nuclear corrections is described.
The results of NNLO QCD analysis of SF are presented  in Section 3.
In Section 4, the $\alpha_S/\pi$-expansion  of the
Gross-Llewellyn Smith sum rule is considered
and expansion parameters are found.

\vskip 1cm

{\section {Method of QCD analysis}}

{\subsection {Jacobi Polynomial Expansion Method}}

 We use, for the QCD analysis,
the Jacobi polynomial expansion method proposed in \cite{PaSu}.
It was developed in \cite{Dubna}
 and applied for the  3--loop  order of perturbative QCD (pQCD)
to fit  $F_2$ \cite{PKK} and $xF_3$ data \cite{KKPS1,KKPS2,Sasha1,Sasha2}.

Following the method \cite{Dubna}, we can write
 the structure  function $~xF_3~$ in the form:
\begin{equation}
xF_{3}^{pQCD}(x,Q^2)=x^{\alpha }(1-x)^{\beta}\sum_{n=0}^{N_{max}}
\Theta_n^{\alpha , \beta }
(x)\sum_{j=0}^{n}c_{j}^{(n)}{(\alpha ,\beta )}
M_{3}^{pQCD} \left (j+2, Q^{2}\right ),   \\
\label{e7}
\end{equation}
where $~\Theta^{\alpha \beta}_{n}(x)~$ is a set of Jacobi polynomials and
$~c^{n}_{j}(\alpha,\beta)~$ are coefficients of the series of
$~\Theta^{\alpha,\beta}_{n}(x)~$ in powers of $x:$
\begin{equation}
\Theta_{n} ^{\alpha , \beta}(x)=
\sum_{j=0}^{n}c_{j}^{(n)}{(\alpha ,\beta )}x^j .
\label{e9}
\end{equation}

The $Q^2$ - evolution of the moments $M_3^{pQCD}(N,Q^2)$ is
given by the well-known perturbative QCD \cite{Buras,Yndur} formula:

\begin{eqnarray}
M_3^{pQCD}(N,Q^2)
& =& \left [ \frac{\alpha _{S}\left ( Q_{0}^{2}\right )}
{\alpha _{S}\left ( Q^{2}\right )}\right ]^{d_{N}}
H_{N}\left (  Q_{0}^{2},Q^{2}\right )
M_3^{pQCD}(N,Q^2_0) ,~~~N = 2,3, ...  \label{m3q2} \\
d_N & = & \gamma^{(0),N}/2\beta_0,
. \nonumber
\end{eqnarray}

 Here ~$\a_s(Q^2)$~ is the next--to--next--to--leading order
strong interaction constant,
~$\gamma^{(0)NS}_{N}$~ are the nonsinglet leading order anomalous
dimensions,
and the factor ~$H_{N}\left (  Q_{0}^{2},Q^{2}\right )$~
contains next-- and  next--to--next--to--leading order QCD
corrections to the coefficient functions and anomalous dimensions
\footnote{For reviews and references on higher order
QCD results, see\cite{vanNeervenHERA}.}
and is constructed in accordance with \cite{KKPS1,KKPS2} based on
theoretical results of \cite{VZ1}--\cite{LRV}.

The unknown coefficients $M_3(N,Q^2_0)$ in (\ref{m3q2}) could be
 parametrised as Mellin moments of some function:
\begin{eqnarray}
M_3^{pQCD}(N,Q^2_0)&=&\int_{0}^{1}dx{x^{N-2}}a_1x^{a_2}(1-x)^{a_3}
(1+a_4 x),
~~~ N = 2,3, ...
\label{Mellf30}
\end{eqnarray}
Here coefficients $a_i$ should be found  by the QCD fit of
experimental data.

The target mass corrections (TMC) are included  into our  fits
through the Nachtmann moments \cite{Nacht} of the SFs:

\begin{eqnarray}
M_{n}^{pQCD}(Q^2)=\int_{0}^{1}dx\xi^{n+1}/(x^3)F_3(x,Q^2)
(1+(n+1)V)/(n+2),
\label{f3}
\end{eqnarray}
where
$\xi=2x/(1+V)$, $V=\sqrt{1+4M_{nucl}^2x^2/Q^2}$ and
$M_{nucl}$ is the mass of a nucleon.
We are taking into account the
order $O(M_{nucl}^2/Q^2)$ corrections:
\begin{eqnarray}
M_{n}(Q^2)&=&M_n^{pQCD}(Q^2)+\frac{n(n+1)}{n+2}\frac{M_{nucl.}^2}{Q^2}
M_{n+2}^{pQCD}(Q^2)
\end{eqnarray}
where $M_{n}$ are the Mellin moments of measured $xF_3$ SF.
\vskip 0.5cm

{\subsection {Higher Twist Correction}}

 We take into account the HT contribution
following the method of \cite{VM} and \cite{Sasha1,Sasha2,KKPS2}.
To extract the HT contribution, the SF is
parametrized as follows:
\begin{eqnarray}
xF_3(x,Q^2)=xF_3^{pQCD}(x,Q^2)+{h(x)\over Q^2},
\label{xf3}
\end{eqnarray}
where the $Q^2$ dependence of the first term in the r.h.s is determined
 by perturbative QCD.

The function $h(x)$  as well as the parameters $a_1-a_4$ and
scale parameter $\Lambda$ should be determined by fitting the
experimental data.

{\subsection {Nuclear Correction}}

We have used the covariant approach in the
light-cone variables \cite{TOK91,BRAUN94} to estimate the
 ratio  of structure functions
\begin{equation}
R_F^{D/N}={F_3^D(x,Q^2)\over F_3^N(x,Q^2)}
\label{eq:r1}
\end{equation}
 and  to perform  the joint NNLO QCD analysis of the data \cite{CCFR97}
on the       structure function $F_3$ (for the NLO analysis,
 see \cite{SID}).

We would like to remind that the ratio $R_F^{A/N}$ describes
the influence of  nuclear medium on the structure of
a free nucleon in the process. We use the approximation that
 $R_F^{D/N}=R_F^{Fe/N}$. It gives lower estimation on the effect
of nucleon Fermi motion in a heavy nucleus.

The  covariant approach in the light-cone variables is based on the
relativistic deuteron wave function (RDWF) with one nucleon
on mass shell. The RDWF  depending on one variable, the virtuality of
nucleon $k^{2}(x, k_{\bot})$,  can be expressed via the $Dpn$ vertex
function ${\Gamma}_{\alpha}(x,k_{\bot})$.
The  deep-inelastic  structure function of neutrino-deuteron scattering
 $F_3^D$ in the approach  can be written as follows

\begin{equation}
F_3^D (\alpha , Q^2) = \int_{\alpha}^{1} dx {\ }{d^2}k_{\bot}
\ \Delta (x,k_{\bot})
\cdot  F_3^N(\alpha /x, Q^2).
\label{eq:r2}
\end{equation}

The nucleon  SF is defined as $F_3^N=(F_3^{\nu N}+F_3^{\bar{\nu} N})/2$,
\ $\alpha = -q^2/2(pq)$.
The function $\Delta (x,k_{\bot})$ describes the left (right)-helicity
distribution  for an active nucleon (antinucleon) that carries away
 the fraction of  deuteron momentum  $x = k_{1+}/p_{+}$
 and  transverse momentum $k_{\bot}$. It is expressed via
the RDWF   $\psi_{\alpha} (k_1)$  as follows

\begin{equation}
\Delta (x,k_{\bot}) \propto Sp\{ \bar {\psi}^{\alpha}(k_1)
\cdot (m+\hat k)\cdot
\psi^{\beta}(k_1)\cdot \hat q \cdot {\sigma}^{\mu \nu} \cdot {\gamma}_5
\cdot\rho_{\alpha \beta}^{(S)}
\cdot {\epsilon}_{\mu \nu \gamma \delta} q^{\gamma} p^{\delta}
  \},
\label{eq:r3}
\end{equation}
where $\rho_{\alpha \beta}^{(S)}$ is the
polarization density matrix for an unpolarized deuteron.
Note that in the approach the use of the distribution function
$\Delta (x,k_{\bot})$ includes not only usual $S$- and $D$-wave
components
of the deuteron but also a $P$-wave component. The latter describes
the contribution of  $N\bar N$-pair production. The contribution of this
mechanism is small over a momentum range ($x < 1$).

The nuclear effect in a deuteron for  the
 $\nu +D\rightarrow {\mu}^{-}+X$   process
 has been estimated in \cite{SID}.
 It has been found that the ratio $R_F^{D/N}$ is practically
independent of the parametrization of parton distributions
\cite{MG92,JGM91,DWD84}  and the nucleon SF \cite{ALK94}
over a wide kinematic range of  $x=10^{-3}-0.7,\
Q^2 = 1 - 500\ (GeV/c)^2$. The curve has an oscillatory feature
and  cross-over point $x_0:  R_F^{D/N}(x_0)=1,\ x_0\simeq 0.03$.

Thus, the obtained results give evidence that the
function $R_F^{D/N}$ is defined by
the structure of the relativistic deuteron wave function and  can
be used to extract the nucleon SF $F_3^N$ from the  experimentally
known deuteron one

\begin{equation}
F_3^N (x,Q^2) =  {[R_F^{D/N}(x)]}^{-1} \cdot F_3^D (x,Q^2).
\label{eq:r4}
\end{equation}

 The performed analysis of the nuclear correction for the nucleon SF
also   allows  one  to consider the influence of the nuclear effect
on the
GLS sum rule \cite{GLS}:

\begin{equation}
S_{GLS}=\int _{0}^{1} F_3^N(x){dx}.
\label{eq:r5}
\end{equation}

We  have used the result on the  $R_F^{D/N}$ ratio
to study the $x$ and $Q^2$-dependences  of the
 GLS integral
\begin{equation}
S_{GLS}(x,Q^2) =\int _{x}^{1}\  F_3^N(y,Q^2) dy.
\label{eq:r6}
\end{equation}

It has been shown in \cite{SID}  that the nuclear effect of
Fermi motion is very important  for the NLO QCD analysis of the
structure function $F_3$ and verification of the
Gross-Llewellyn Smith  sum rule
over a wide region of $Q^2=3-500\ (GeV/c)^2$. Therefore in the paper, we
take into account  the nuclear effect within the joint NNLO QCD analysis
of the structure function and  of the Gross-Llewellyn Smith  sum rule.

\vskip 1cm

{\section { NNLO QCD analysis of Structure function $xF_3^N$}}

In this section, we perform the QCD analysis of the $xF_3^N$
experimental
 data \cite{CCFR97}
taking into account the 3-loop QCD, higher twist and nuclear
corrections.
We  consider as a first approximation
that $R_F^{Fe/N}=R_F^{D/N}\equiv R$.

\begin{equation}
xF_3^N(x,Q^2)= a_1(Q^2)x^{a_2(Q^2)}
(1-x)^{a_3(Q^2)}(1+a_4(Q^2)x) +{h(x)\over Q^2},
\label{eq:r7}
\end{equation}
Constants $h(x_i)$ (one per $x$--bin) parametrize the HT $x$-dependence.
The points ${x_i}$ are chosen in accordance with \cite{CCFR97}:
0.0075 -- 0.75.

The values of  the parameters $a_i(Q^2_0)\ (i=1-4)$ ,
scale parameter $\Lambda$  and
constants $h(x_i)$ have been  determined by fitting the
whole  set of $xF_3$ data  \cite{CCFR97} (116 experimental points)
for different values of $Q^2_0$ in the
kinematic region: $2~GeV^2\leq~Q^2~\leq~200~GeV^2$.
{\footnote{The number of flavors is taken to be equal to 4.}}
Only statistical errors are taken into account. The $\chi ^2$
parameter is found to be about 105 for 116 experimental points.

Figure 1(a-d) shows the parameters SF $a_i$ at the points $Q^2=$
1.5 -- 200\ $(GeV/c)^2.$ The $\Lambda_{{\overline{MS}}}$ parameter
was found in the
interval $(210-250)~MeV$ with statistical error about $\pm 20~MeV$.

Figure 2 demonstrates the $x$-dependence of the HT
contribution $h(x_i)$.

We use the result of our NNLO QCD analysis presented in figs. 1 and 2
to obtain  the parametrisation of $xF_3$ SF in form (15).

The $Q^2$-dependence of the coefficients $a_i$
is parametrized as follows:

\begin{equation}
a_i(Q^2)=\sum_{j=0}^2 c_j\cdot z^j,\ \ z=\log(Q^2)
\label{eq:r8}
\end{equation}
and is shown in Fig.1 by solid lines.
The coefficients $c_i$  are presented in Table 1.

\vskip 0.5cm

{{\bf Table 1.}
 Coefficients ${c_i}$ for the parametrization of the function
  $xF_3(x,Q^2)$}

\vskip 0.5cm
\begin{center}
\begin{tabular}{|c|c|c|c|c|} \hline
\baselineskip =3.\baselineskip
         & $a_1$    &$a_2$        &$a_3$       &$a_4$
               \\  \hline
$c_{0} $ & $\ 3.1430 $  &$\ 6.6704e-1$  &$\ 3.2753e+0$
 &$\ 2.92880e+0$    \\
$c_{1} $ & $\ 1.6794 $  &$\ 4.0720e-3$  &$\ 8.2294e-1$
  &$ -1.65755e+0$   \\
$c_{2} $ & $ -0.3896 $  &$ -7.7966e-3$  &$ -1.3154e-1$
  &$\ 3.09344e-1$   \\ \hline
\end{tabular}
\end{center}

\vskip 1.5cm

Figure 1 shows the dependence of coefficients $a_i$ on $Q^2$.
The black circles are experimental points, lines are results of the
fit.

The  parameter $a_2$ slowly depends on $Q^2$ and corresponds
to the theoretical estimation $a_2^{theor}\simeq0.68$
\cite{Manaen}.

Our results for   $a_3(Q^2)$ could be compared to the well known
Buras-Gaemers parametrization \cite{BG}:

\begin{equation}
 a_3=\eta _1 + \eta _2\cdot
 \frac{ln(Q^2/\Lambda ^2)}{ln(Q^2_0/\Lambda ^2)}
\label{eq:r9}
\end{equation}
At the point $Q^2_0=1.8~GeV^2$ it gives \cite{Buras}
$\eta_1=2.6$          and $\eta_2=0.8$.        Our result is
$\eta_1=3.48\pm0.03$  and $\eta_2=1.18\pm0.09$. It is not far
from the pQCD theoretical estimation  \cite{Korch} based
on the  relation
\begin{equation}
\frac{d}{dln(Q^2)} a_3(Q^2)=
\frac{8}{3\pi}\alpha_S(Q^2)+O(\alpha_S^2(Q^2))
\label{eq:r10}
\end{equation}
and gives $\eta_2^{theor}=0.64$.

The term describing the higher twist correction is written in the form
\begin{equation}
h(x)= \sum_{i=0}^3  d_i\cdot z^i,\ \ z=\log(x)
\label{eq:r11}
\end{equation}
 and is shown in Fig.2 . The values of coefficients ${d_i}$ are
 presented
 in Table 2.
\vskip 0.5cm

{{\bf Table 2. }
 Coefficients ${d_i}$ for the parametrization of the function  $h(x)$}
\vskip 0.5cm

\begin{center}
\begin{tabular}{|c|c|c|c|} \hline
\baselineskip =3.\baselineskip
 $d_0$      &$ d_1 $     &$d_2$  & $d_3$   \\  \hline
  0.2329   & 0.8060          & 0.7308      &0.1842    \\ \hline
\end{tabular}
\end{center}


Thus, the NNLO QCD analysis of experimental data  of the
structure function $xF_3$ has been performed  and the  parametrizations
of the NNLO perturbative QCD, TMC, nuclear effect and higher twist
corrections  have been obtained.

\vskip 1cm

{\section {
Experimental constraints on coefficients of $\alpha_S$-expansion
of  Gross-Llewellyn Smith sum rule}}

In this section, we would like to show the status
of the NNLO QCD analysis of the experimental data \cite{CCFR97}.

The $Q^2$--dependence of the  parameters $a_i(Q^2)$ allows us to
study the behavior of $S_{GLS}(Q^2)$ in a wide kinematic region
of the momentum transfer squared.

Figure 3 shows the  dependence of $S_{GLS}(Q^2)$ on
$\alpha_S/\pi$ in thw 3-loop approximation.
The black and open circles correspond
to the NNLO QCD analysis with and without
nuclear corrections, respectively. The statistical errors are about
$\pm 0.4$
and are not presented at the figure.

Figure 3  demonstrates the increase of $S_{GLS}(Q^2)$  with decreasing
$\alpha_S/\pi$. The result is in qualitative agreement with
the $Q^2$-dependence of the sum rule found in \cite{KS} in the
NLO QCD analysis
without target  mass corrections, higher twist and nuclear effect taking.
A  similar tendency  in the NLO QCD approximation
with target  mass corrections and nuclear effect was found
in \cite{SID}.
For the estimation of the order $O(\alpha_s^4)$
and power corrections, see \cite{ckks}.
Figure 3 shows a considerable  sensitivity of
$S_{GLS}(Q^2)$  to the nuclear correction.

The pQCD predictions for $\alpha_S/\pi$-expansion  of the
Gross-Llewellyn Smith sum rule  up  to $(\alpha_S/\pi)^2$   could be
presented in the form

\begin{equation}
S_{GLS}^{\it theor}(Q^2)=c_0\cdot (1 +c_1\cdot (\alpha_S/\pi)
 + c_2\cdot (\alpha_S/\pi)^2).
\label{eq:r12}
\end{equation}
The coefficients $c_1$  and $c_2$ have been calculated
in \cite{gls3loop}:
$c_1= -1, \ \ c_2= -3.25 $.

The $\alpha_S$-dependence of $S_{GLS}$ presented
in Fig.3  could be parametrized by the parabola:
\begin{equation}
S_{GLS}^{\it exp}(Q^2)=
s_0+s_1\cdot (\alpha_S/\pi) + s_2\cdot (\alpha_S/\pi)^2.
\label{eq:r13}
\end{equation}
This expansion allows us to directly compare the results of
experimental data analysis with NNLO QCD calculations.
The values of $s_0, s_1, s_2$
for the interval $\alpha_S/\pi < 0.09$
are presented in Table 3.

\vskip 0.5cm

{{\bf Table 3.}  The  coefficients of the
  Gross-Llewellyn Smith  integral
$S_{GLS}(Q^2)$
expansion in $\alpha_S/\pi$.}\\

\begin{center}
\begin{tabular}{|c|c|c|c|} \hline
\baselineskip =3.\baselineskip
           &NNLO+HT         &NNLO+HT+NC   &Theor. \cite{gls3loop}
            \\  \hline
 $ s_0$      &$ 2.86\pm 0.01$  &$ 2.74\pm 0.01$      &$\ 3.00$
    \\
 $ s_1$      &$-4.93\pm 0.32$  &$-2.22\pm 0.23$      &$ -3.00$
  \\
 $ s_2$      &$ 0.98\pm 2.54$  &$-7.86\pm 1.74$      &$-9.75 $
 \\ \hline
\end{tabular}
\end{center}

From Table 3 one can see a high sensitivity of parameters
$s_i$ to the nuclear correction.
The obtained results for the coefficients $s_i$ show that the
consideration of the nuclear correction gives good agreement
with the theoretical calculation  of next-to-leading and
next--to-next-leading
order QCD  corrections.

\vskip 0.5cm

{\section {Conclusion}}

The detailed NNLO QCD  analysis of the         structure function $xF_3$
of new CCFR  data
including  the target  mass, higher twist and nuclear
corrections  was performed.
 The parametrizations of perturbative and power
terms  of the structure function were constructed.
The results of  QCD analysis of the structure function  were used to
study the $Q^2$-dependence of the Gross-Llewellyn Smith sum rule.
The $\alpha_S/\pi$-expansion  of $S_{GLS}(Q^2)$  was exemined
and expansion parameters $s_1, s_2, s_3$ were found.
We would like to emphasize that the consideration of the nuclear
correction allows us to achieve a good qualitative agreement between the
results obtained  from the experimental data  by the NNLO QCD analysis
and perturbative QCD predictions  for the
Gross-Llewellyn Smith sum rule in
next-to-leading and next-to-next-to-leading order.

\begin{center}
{\large \bf Acknowledgement}
\end{center}
This work was partially supported by Grants of the Russian
Foundation for Fundamental Research under   No. 95-02-05061 and No.
95-02-04314 and INTAS project N 93-1180.

\newpage
\begin{minipage}{4cm}

\end{minipage}
\vskip 3cm

\begin{center}
{\bf  FIGURE CAPTIONS}\\[1.cm]
\end{center}
{\bf Figure 1.} Dependence of the coefficients $a_1$-$a_4$ on $Q^2$.
The black circles are experimental points,
lines are results of the fit.\\[0.5cm]
{\bf Figure 2.}
The function $h(x)$ vs $log(x)$ describing a higher twist correction.
The black circles are experimental points,
the line is the result of the fit.\\ [0.5cm]
{\bf Figure 3.}
The $\alpha_S/\pi$-expansion of the Gross-Llewellyn Smith sum rule
$S_{GLS}(Q^2)$.
Black and open circles are results of NNLO QCD analysis obtained
with and without nuclear correction, respectively.\\
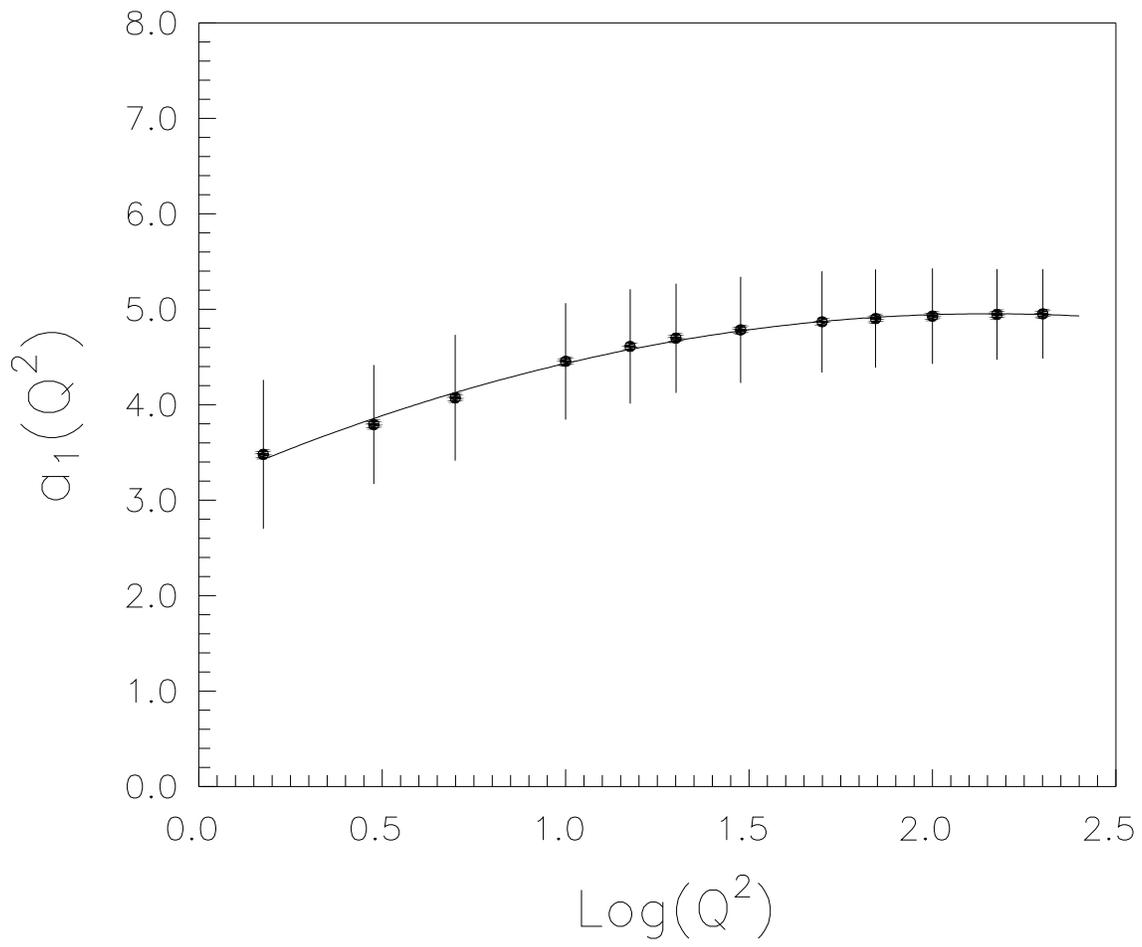
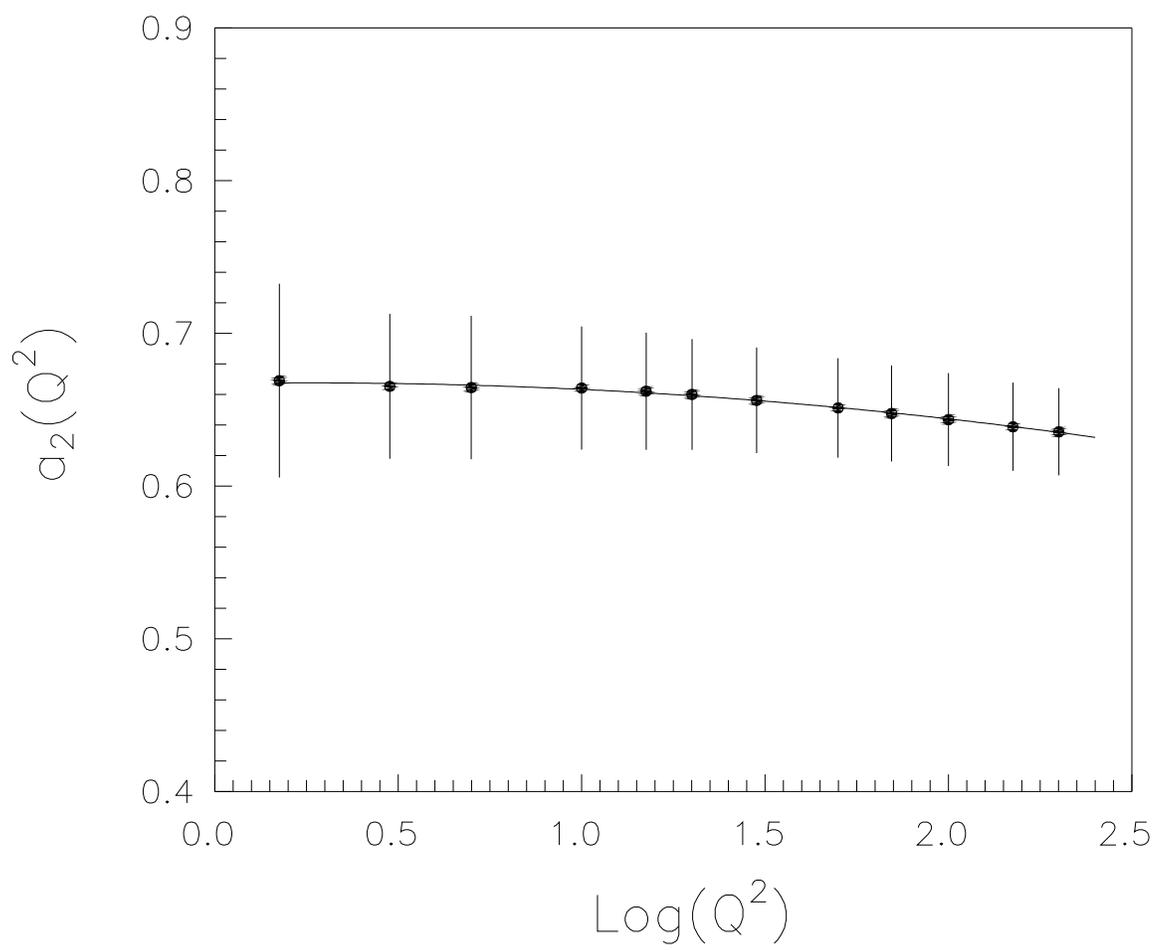
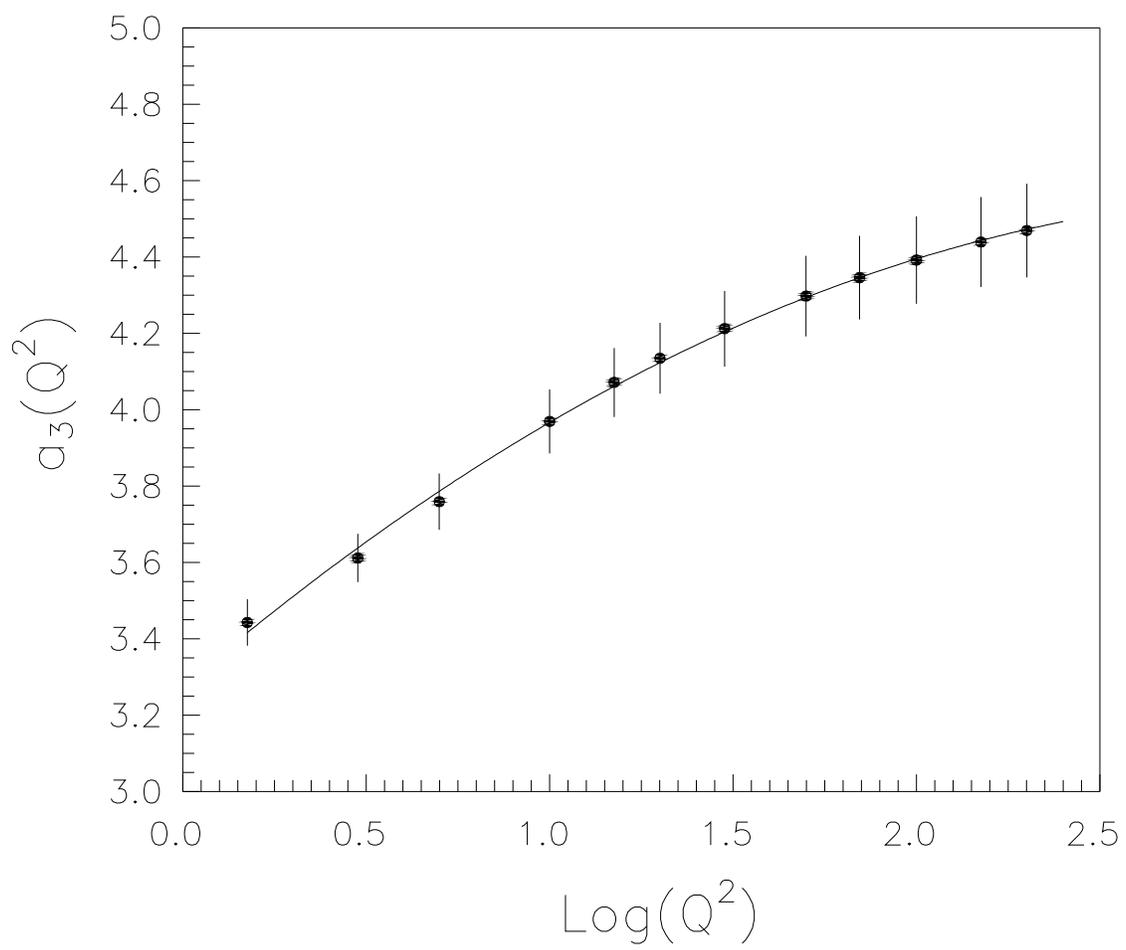
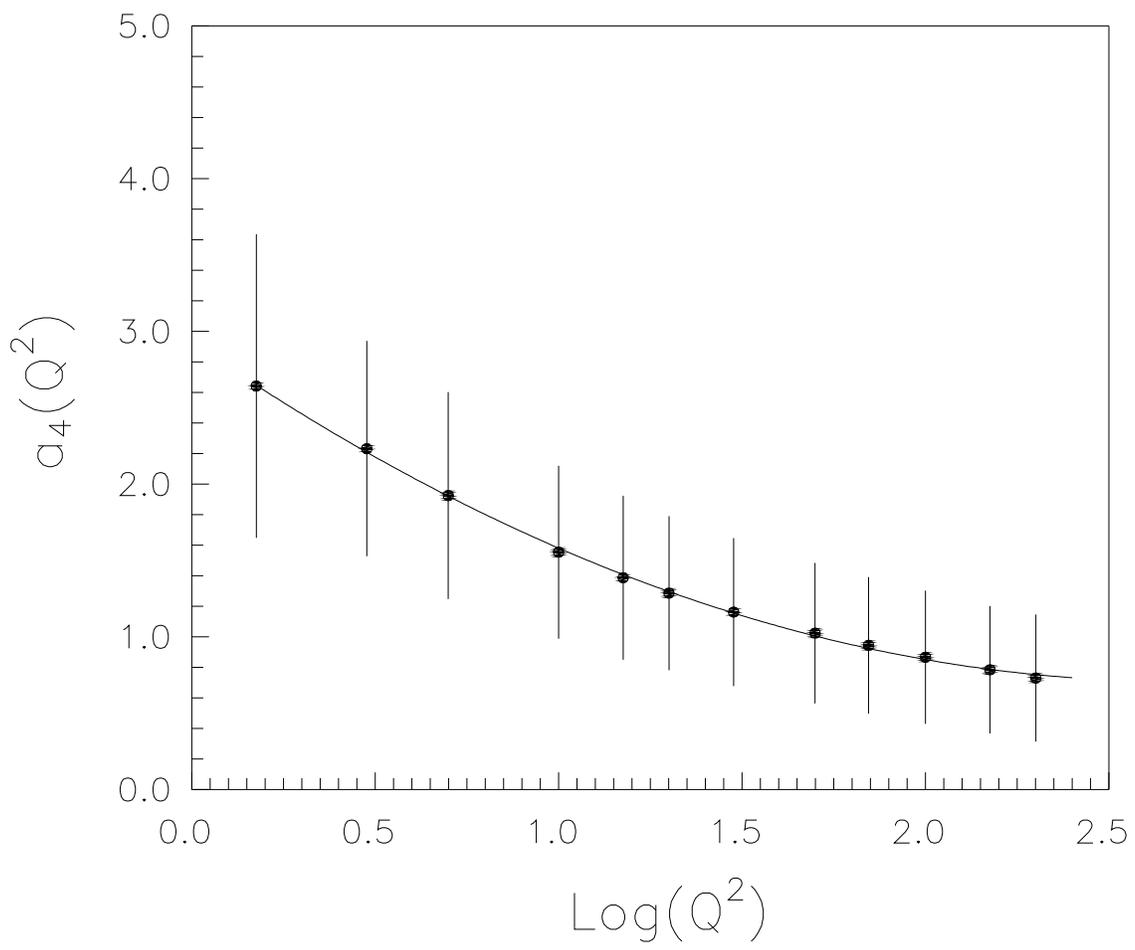
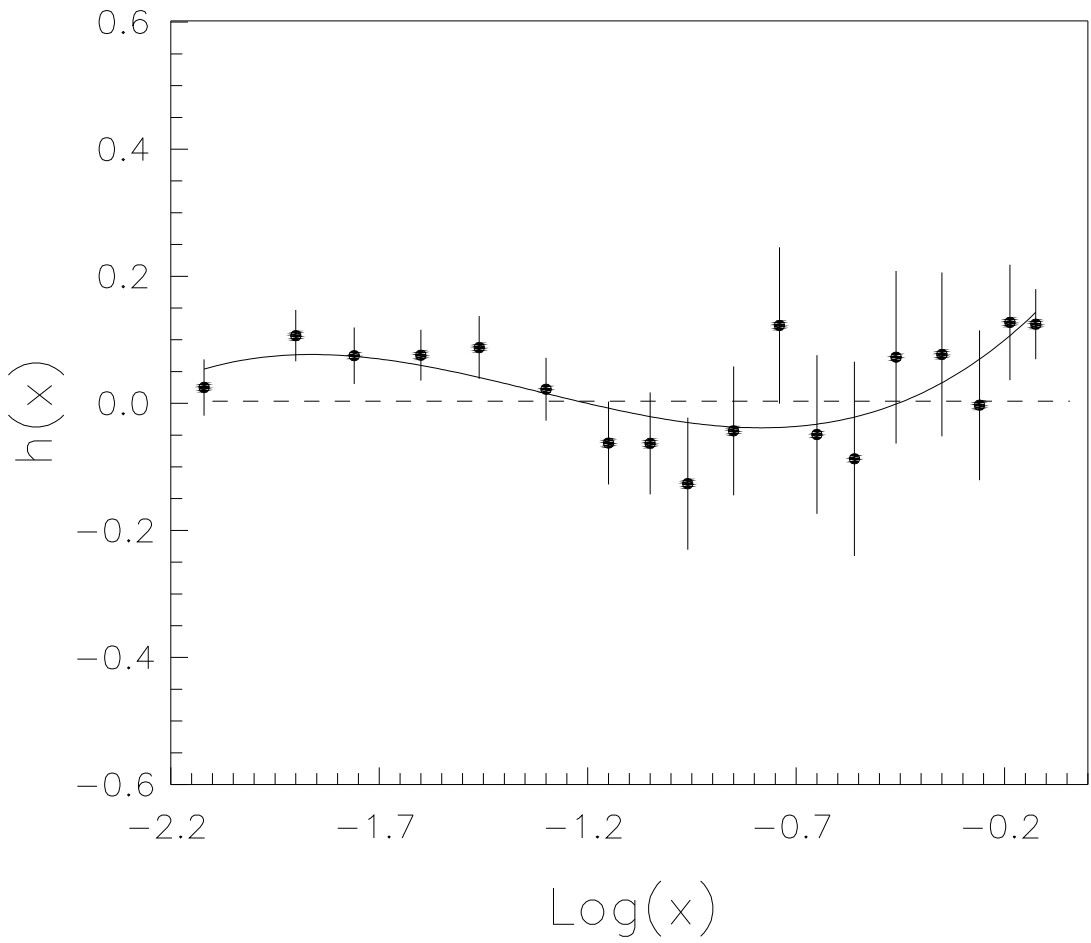
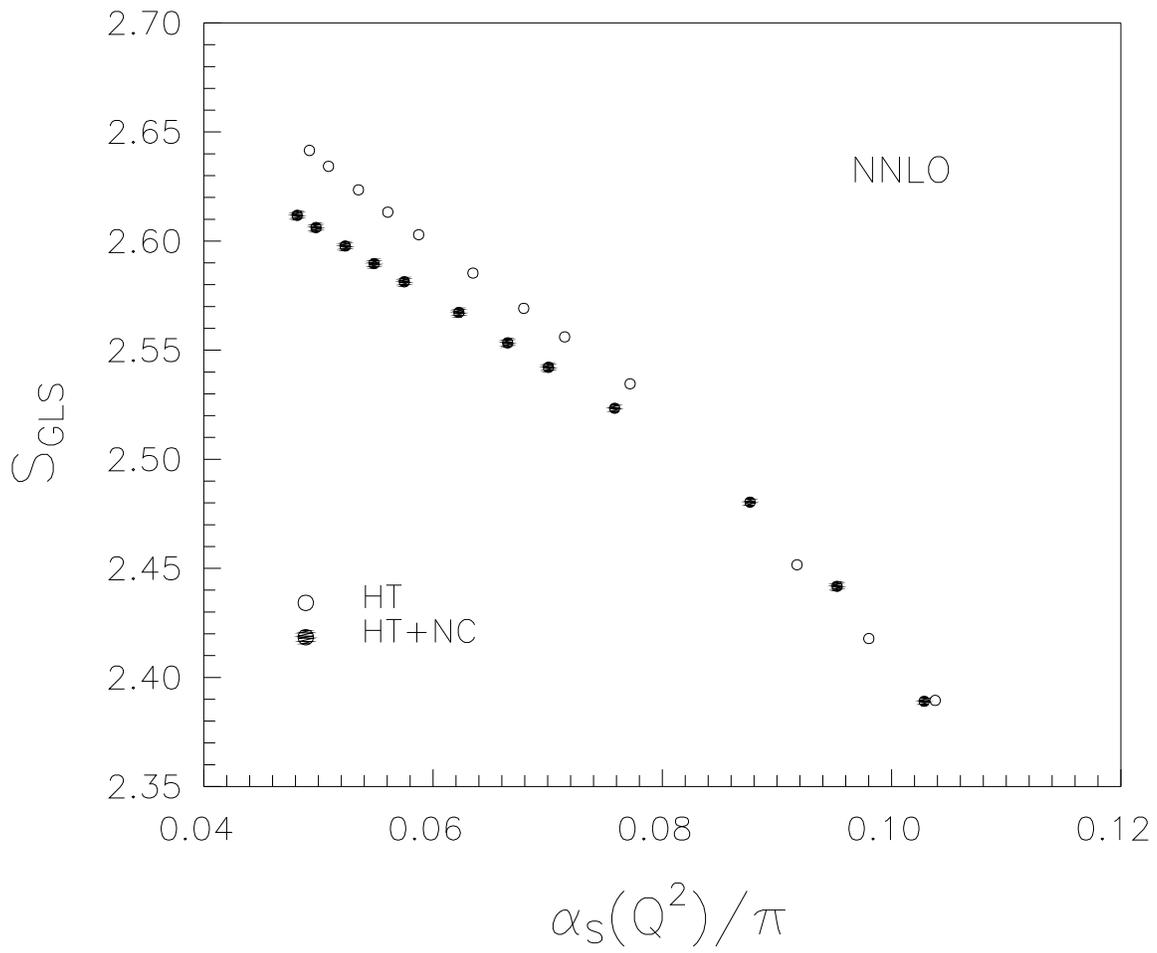
\end{document}